# DYNAMIC APERTURE CALCULATION FOR THE DAΦNE-II PROJECT

E. Levichev, P. Piminov[#],
BINP, Lavrentiev 13, Novosibirsk 630090, Russia

## INTRODUCTION

The *DAΦNE-II* project is considered now as a possible candidate for upgrading the *DAΦNE* electron-positron collider and improving its luminosity. The basic idea is to use a strong RF focusing mechanism [1] to compress a bunch at the interaction point and hence to get a chance for the vertical beta reduction.

An intrinsic feature of the strong RF focusing is a large value of synchrotron tune and one can assume intuitively that just like the betatron strong focusing results in the transverse dynamic aperture limitation, the synchrotron strong focusing can provide the same but for the energy dependent dynamic aperture.

In order to check that, we were performing a 3-D simulation of the *DAΦNE-II* dynamic aperture under various assumptions (weak and strong RF focusing). The results of this tracking are presented in the paper.

## DAΦNE-II LATTICE

The *DAΦNE-II* lattice proposed by C. Biscari [2] has rather large negative momentum compaction factor necessary for effective bunch length squeezing and as a promising counteraction against the microwave bunch lengthening [3]. In order to achieve it, the arc cell contains negative and positive curvature dipole magnets, a number of quadrupole magnets to focus the beam in the transverse plane and sextupole magnets to compensate for rather large natural chromaticity.

The betatron and dispersion functions of the *DAΦNE-II* are shown in Fig.1.1 and Fig.1.2 while the main parameters are listed in Table 1.1. Dispersion-free straight sections are discussed for detector and RF cavities accommodation.

Table 1.1: *DAFNE-2* main parameters.

| Beam energy, $E$ | 511 MeV |
|---|---|
| Circumference, $L$ | 103.45 m |
| Revolution frequency, $f_0$ | 2.898 MHz |
| Revolution period, $T$ | 0.345 μs |
| Betatron tune, $\nu_x/\nu_z$ | 8.792/7.893 |
| Natural chromaticity, $\xi_x/\xi_z$ | -18.4/-37.2 |
| Momentum compaction factor, $\alpha$ | -0.214 |
| Beta-functions at IP, $\beta_x^*/\beta_z^*$ | 50.0/0.25 cm |
| Energy loss per turn, $U_0$ | 35.45 KeV |
| Partition numbers, $J_x/J_z/J_s$ | 1.67/1.0/1.33 |
| Damping times, $\tau_x/\tau_z/\tau_s$ | 5.9/9.9/7.5 ms |
| Horizontal emittance, $\varepsilon_x$ | $6.38 \cdot 10^{-8}$ m·rad |
| Energy spread, $\sigma_{\Delta E/E}$ | $5.53 \cdot 10^{-4}$ |

___________________
[#]piminov@inp.nsk.su

To study the influence of synchro-betatron resonances in the case of the strong longitudinal focusing, the simulation of the non-linear beam behavior was performed with the help of the **ACCELERATICUM** computer code [4].

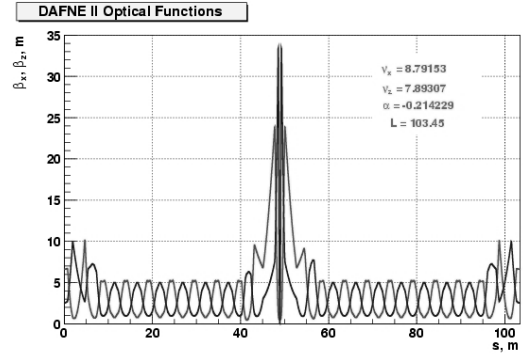

Fig.1.1 *DAΦNE-II* betatron functions.

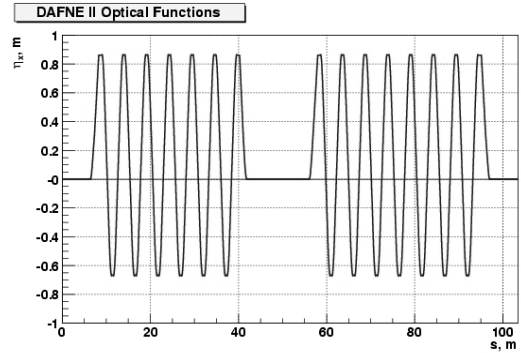

Fig.1.2 *DAΦNE-II* dispersion function.

## SIMULATION RESULTS

The **ACCELERATICUM** code is a general-purpose code to study different aspects of particle motion in a circular accelerator.

It provides a symplectic 6D tracking for the transversely and longitudinally coupled magnetic lattice according to the formalism proposed by G. Ripken in [5]. The formalism uses the canonical variables, which are commonly used in the six-dimensional linear theory

$$\left( x, p_x, y, p_y, \sigma(s) = s - c \cdot t(s), \delta = \frac{E - E_0}{E_0} \right),$$

and which are also canonical in the non-linear formalism if the transformation through the nonlinear elements is performed with the help of Hamiltonian generating functions approach.

Besides the nonlinear dynamics, the 6D tracking allows us to investigate linear parameters of the machine as a function of the beam momentum deviation (betatron functions, dispersion, etc.). For instance, Fig.2.1 shows the nonlinear part of the chromaticity when the linear part is corrected to zero by the sextupoles and Fig.2.2 shows the momentum compaction factor as a function of energy deviation.

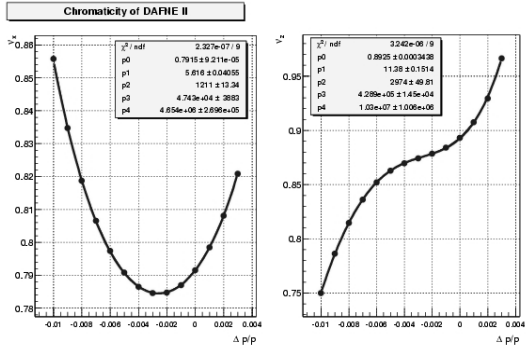

Fig.2.1 Residual tune chromaticity after sextupole corection.

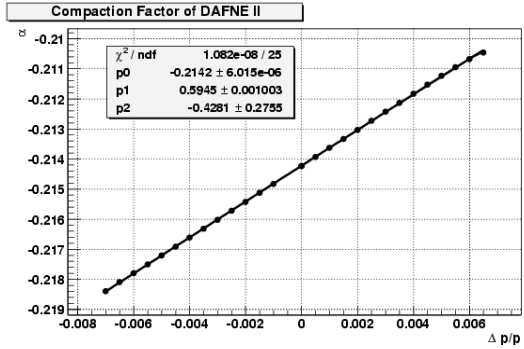

Fig.2.2 Momentum compaction factor vs. energy deviation.

Two families of sextupole magnets (see their integrated strength in Table 2.1) in the arc cells were set to adjust the natural chromaticity to zero.

Table 2.1 Sextupole magnets' integrated strength.

| Name | $(ml)$, m$^{-2}$ |
|---|---|
| SD | -6.06 |
| SF | 2.69 |

Table 2.2 RF parameters of the $DA\Phi NE$-$II$.

| Harmonic number, $h$ | 175 | | |
|---|---|---|---|
| RF frequency, $f_{RF}$ | 507.140 MHz | | |
| RF voltage, $U_{RF}$ | 300 kV | 3 MV | 5.8 MV |
| Synch. phase, $\phi_{RF}$ | 118.45 mrad | 11.818 mrad | 6.112 mrad |
| Synchrotron tune, $\nu_s$ | 0.059 | 0.200 | 0.305 |
| Synch. frequency, $\Omega_s$ | 171.91 kHz | 507.14 kHz | 882.92 kHz |
| Length of bunch, $L_b$ | 20.66 cm | 6.12 cm | 4.02 cm |
| RF bucket width, $P_{max}$ | $2.878 \cdot 10^{-3}$ | $1.057 \cdot 10^{-2}$ | $1.618 \cdot 10^{-2}$ |
| RF bucket length, $L_{max}$ | 28.44 cm | 29.45 cm | 29.50 cm |

To consider dynamic aperture limitation due to the strong RF focusing we have used several values of RF voltage during simulation (see Table 2.2). The case of $U_{RF}$=300 kV corresponds to the weak focusing while 3 MV and 5.8 MV provide strong synchrotron focusing of the bunch at the IP.

## DYNAMIC APERTURE SIMULATION

All plots of the dynamic aperture are presented for the interaction point. We use the 4D (without synchrotron motion) dynamic aperture shown in Fig.3.1 as a reference. The picture is typical for the coupling resonance limitation of a stable area: large 1D aperture (along the $x$-axis) is reduced if the vertical motion with arbitrary small amplitude is switched on.

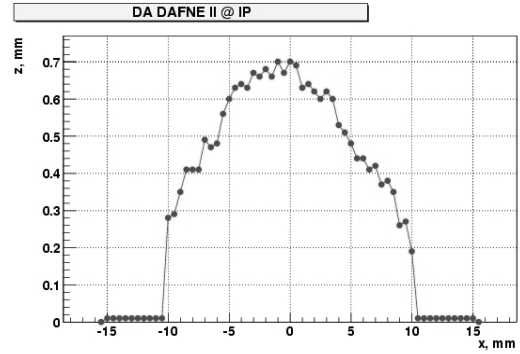

Fig.3.1 4D dynamic aperture of $DA\Phi NE$-$II$ (1000 turns).

Fig.3.2 shows the $DA\Phi NE$-$II$ off-energy dynamic aperture for the constant energy deviation (no synchrotron oscillation is turned on) for the weak (left plot) and strong (right plot) RF focusing. In this case, the limitation of the particle stable area can be explained by different on- and off-energy particle trajectories in the magnetic field.

A rather different situation can be seen in Fig.3.3 and Fig.3.4, where the synchrotron oscillation is taken into account. While for the weak RF focusing the dynamic aperture does not differ much from that with a constant energy deviation (Fig.3.2, left), for the strong RF focusing the dynamic aperture became very small even for $\Delta p/p=0$.

The following schematic mechanism can be proposed. For the general tune-amplitude dependence expressions

$$\Delta\nu_x = C_{xx}A_x^2 + C_{xz}A_z^2,$$
$$\Delta\nu_z = C_{zx}A_x^2 + C_{zz}A_z^2, \quad (3.1)$$

and in the presence of synchrotron oscillation, the resonant condition has the form

$$m_x\left(\nu_{x0} + C_{xx}A_x^2 + C_{xz}A_z^2\right)+$$
$$+ m_z\left(\nu_{z0} + C_{zx}A_x^2 + C_{zz}A_z^2\right) + k\nu_s = n, \quad (3.2)$$

where $\nu_{x0}$, $\nu_{z0}$ and $\nu_s$ are the linear (non-perturbed) tunes for three oscillation modes. Now suppose that at every point of the dynamic aperture curve we have some particular betatron resonance that limits the stable area in this point. Then the synchrotron motion generates a set of satellite resonances, which are represented by lines at the amplitude plane $A_z(A_x)$. The resonance line equation can be defined from (3.2). For the sake of simplicity, consider only the horizontal resonance and main (strongest $k$=1) satellite $m_x(\nu_{x0} + C_{xx}A_x^2) + \nu_s = n$, the following expression for the horizontal position of the satellite resonance line can be deduced:

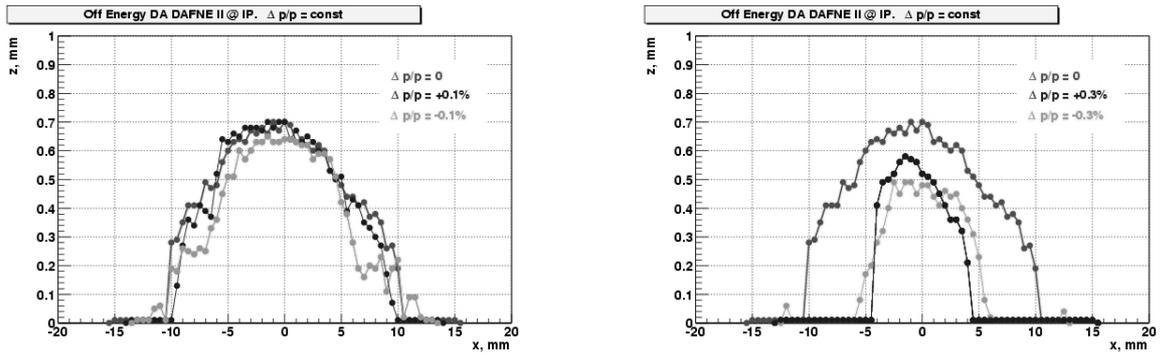

Fig.3.2 Off-energy DA with constant energy deviation (no synchrotron oscillation). $U_{RF} = 300$ kV (left), = 3 MV (right).

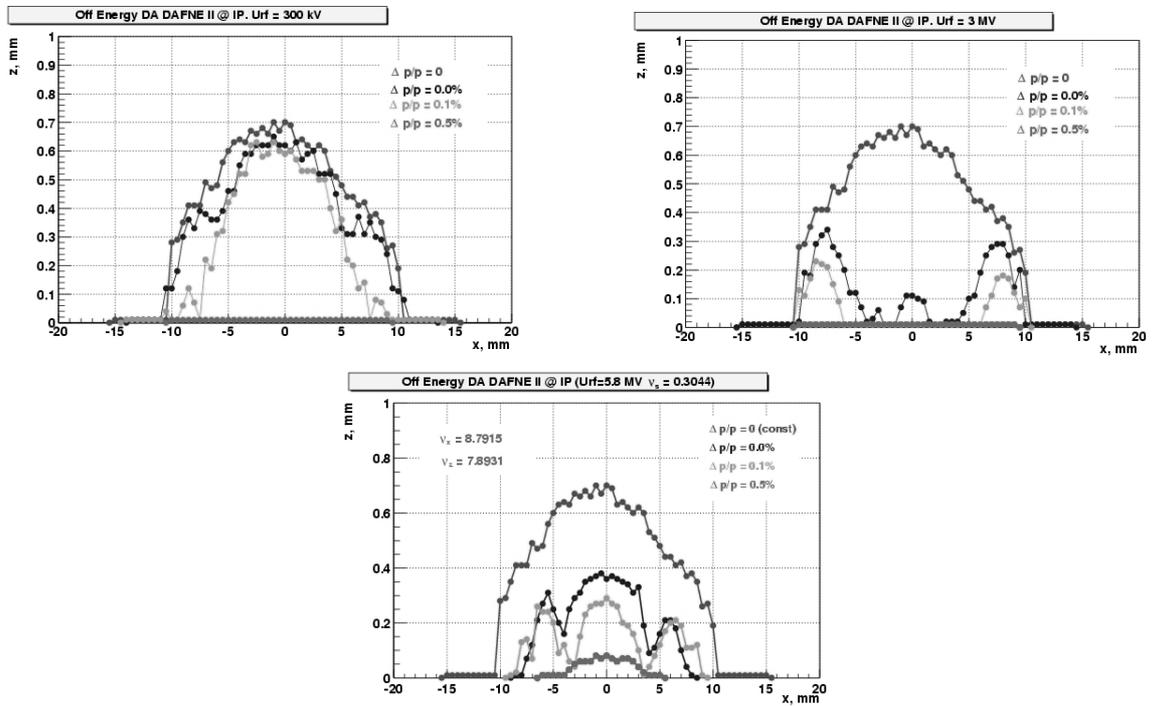

Fig.3.3 Off-energy DA with synchrotron oscillation. $U_{RF}$=300 kV (left plot), = 3 MV (right plot), = 5.8 MV (below).

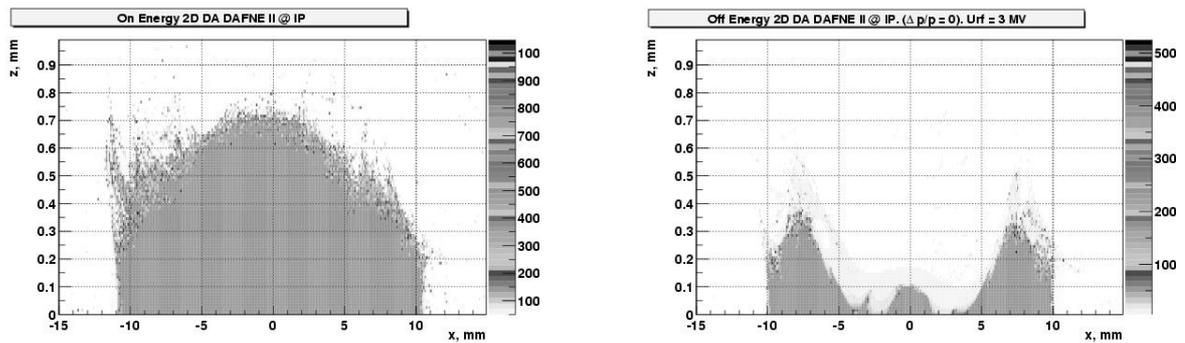

Fig.3.4 The same as in Fig.3.3. High resolution survival plot.

$$A_x = A_{x0}\sqrt{1 - \frac{\nu_s}{m_s \delta}}, \qquad (3.3)$$

where $\delta = \nu_{x0} - n/m$ is a distance from the resonance and $A_{x0} = \sqrt{\delta/C_{xx}}$ is the position of the original ($\nu_s=0$) betatron resonance on the amplitude plane. The strong satellite resonance inside the initial dynamic aperture provides an additional reduction of the stable area as is clearly seen in the right of Fig.3.3 and Fig.3.4.

In the case of weak RF focusing, the situation is not so serious for two possible reasons:
(1) The distance (in the amplitude space) between the main and satellite resonances is small and only slightly distorts later the edge of the dynamic aperture. Some evidence of this fact one can see in the left-hand side of Fig.3.3 (the blue and green curve).
(2) The amplitude of the satellite resonance depends on the synchrotron tune and drops down with the satellite number $k$ (most probably like the Bessel function).

## CONCLUSIONS

The 6D tracking with synchrotron oscillation shows that in the case of strong RF focusing the dynamic aperture of the *DAΦNE-II* is reduced as compared to the week focusing case or constant energy deviation. A possible mechanism of this reduction is that the synchrotron motion produces satellites of the strong sextupole resonances ,which limit the dynamic aperture in the 4D case. The satellite resonances locate inside the initially stable area and additionally reduce it. The following plan for the further study and recover of this phenomenon can be proposed:
(1) More detailed investigation of the satellite behaviour for the weak, strong and intermediate RF focusing, including the satellites amplitude values.
(2) In the case of the strong RF focusing dependence of dynamic aperture on the tune point is to be explored (in other words, more accurate choosing of the betatron and synchrotron tunes). It seems that all the three tunes are important now.
(3) As the satellites resonances location depends on the detuning coefficients, it is necessary to check if it possible to control it by octupole magnets.

## REFERENCES

[1] A.Gallo, P.Raimondi and M.Zobov. Strong RF focusing for luminosity increase. DAFNE Technical Note G-60, Frascati, August 18, 2003.
[2] C.Biscari, Lattice for Longitudinal Low-Beta, these proceedings, 2003.
[3] S.X.Fang, K.Oide, K.Yokoya, et al. Microwave instability in electron rings with negative momentum compaction factor. KEK Preprint 94-190.
[4] Tracking code ACCELERATICUM, VEPP-4M Internal Note, BINP, Novosibirsk, 2003.
[5] G.Ripken – DESY Report 85-084, August 1985.